\documentclass{appolb}

\usepackage{graphicx}
\usepackage{subfigure}
\usepackage{cite}
\usepackage{epsfig}
\usepackage{amssymb}
\usepackage{amsmath,amsfonts}
\usepackage{stackrel}
\usepackage{soul} 
\usepackage{color}

\newcommand{\modifs}[2]{\textcolor{red}{#1}\ \st{#2}}

\newcommand{\ignore}[1]{}

\begin{document}
\title{The X(3872) as a mass distribution\thanks{Talk by E. Ruiz
    Arriola at ``Excited QCD 2020'', Krynica Zdr\'oj, Poland, February
    2-8, 2020.}  \thanks{Work supported by the Spanish MINECO and
    European FEDER funds (FIS2017-85053-C2-1-P, FPA2016-77177-C2-2-P), Junta de
    Andaluc\'{\i}a (grant FQM-225). This publication is supported by EU Horizon 2020 research and innovation programme, STRONG-2020 project, under grant agreement No 824093}
} \author{E.~Ruiz Arriola$^{\, a}$, P. Garc\'{\i}a Ortega~$^{\, b}$
  \address{$^{a}$Departamento de F\'{\i}sica At\'omica, Molecular y
    Nuclear and Instituto Carlos I de F\'{\i}sica Te\'orica y
    Computacional, Universidad de Granada, \\ Avenida de Fuente Nueva
    s/n, 18071 Granada, Spain } \address{$^{b}$Departamento de F\'\i
    sica Fundamental and Instituto Universitario de F\'\i sica Fundamental y Matem\'aticas (IUFFyM), Universidad de Salamanca, \\ E-37008 Salamanca, Spain} \\ } \maketitle
\begin{abstract}
All existing experimental evidence of the bound state nature of the
X(3872) relies on considering its decay products with a finite
experimental spectral mass resolution which is typically $\Delta m \ge
2$ MeV and much larger than its alleged binding energy, $B_X
=0.00(18)$MeV. On the other hand, there is a neat cancellation in the
$1^{++} $ channel for the invariant $D {\bar D}^*$ mass around the
threshold between the continuum and bound state contribution. We
discuss the impact of this effect for X(3872) at finite temperature,
in prompt production in pp collisions data with a finite pT or the
lineshapes of specific production experiments of exotic states
involving triangle singularities
\end{abstract}
\PACS{11.10.Wx, 12.38.-t, 12.38.Lg}

\section{Introduction}

While the QCD spectrum is expected to describe the experimentally
observed hadronic states, the precise manner how this is supossed to
happen is not at all clear in the real world where most states belong
to the continuum. Besides, a quantitative measure of the ``distance''
between two different spectra is never used. Moreover, with the
exception of low-lying hadronic states, which are truely stable
particles at the strong interactions level, most of the remaining
reported states are resonances with a finite lifetime and hence
undergo strong decays. Besides, the very definition of the hadronic
density of states includes the corresponding resonance background,
which has no universal definition and its phenomenological
determination depends on the particular process where the resonance
shows up.

Two extremely different and complementary alternatives to this issue
rely on either the individual level by level analysis based on
hadronic reactions or on the collective approach addressing bulk
properties such as level density of states or thermodynamic
properties. While the Particle Data Booklet~\cite{Tanabashi:2018oca}
summarizes all relevant information concerning established states with
a given confidence level, there always arises the question which are
the {\it a priori } pre-requisites making a possible observed peak or
bump in a hadronic reaction qualify as an eligible hadronic state, so
that a majority vote of the experts prevails. The thermodynamic
approach does offer a global perspective based on the Hadron Resonance
gas which works fairly well in the confined phase of a hot hadronic
medium and has been checked both in ultrarrelativistic heavy ions
collisions as well as on lattice QCD.  At the present moment, and
after some long discussions on the nature of states it is fair to say
that the currently accepted PDG states are relatives to the naive
quark model, either below or above the different two-heavy-light meson
states, such as $D \bar D$, $D \bar D^*$ etc. This also suggests that
the Hadron Resonance Gas corresponds to blindly implemented {\it all}
the listed PDG states assuming their existence is established (see
e.g. Ref.~\cite{Arriola:2014bfa} for a pedagogical presentation and
references therein.)

The proliferation of X,Y,Z states in the heavy sector above the
charmonium in the last decade at different experimental facilities
makes this pertinent question more accute in view of the fact that
many of them are of molecular nature and hence weakly bound: should
these states be reported as genuine contributions to the hadronic
level density ? In this talk we share our views on the subject
taking the prominent  $X(3872)$ state as an example in the $J^{PC}=1^{++}$
channel~\cite{Ortega:2017hpw,Ortega:2017shf,Ortega:2019fme}. 

The question was raised long ago by Dashen and
Kane~\cite{Dashen:1974ns} where they pointed out that the mere
counting of hadrons depends on the energy or mass resolution $\Delta
M$ with which we decide to bin the states. If we take $\Delta M$ to be
the typical $SU(3)$-splitting some states may not count, due to a
cancellation of the bound state and the continuum contributions in the
level density.  As an example they provided the uncontroversial
deuteron state, a $J^{PC}=1^{++}$ weakly bound state of a proton and a
neutron with a total mass of $ M_d = 2.013$ GeV, where the binding
energy $B=2.2$ MeV is about one per mil the total mass and their mean
separacion is much larger than the size of the nucleons (see also
Ref.~\cite{Arriola:2015gra}).  According to this observation the
deuteron should not be listed in the PDG. Thus, what can be said in
this regard about the $X(3872)$ ?

\begin{figure}[h]
  \begin{center}  
  \includegraphics[width=.45\textwidth]{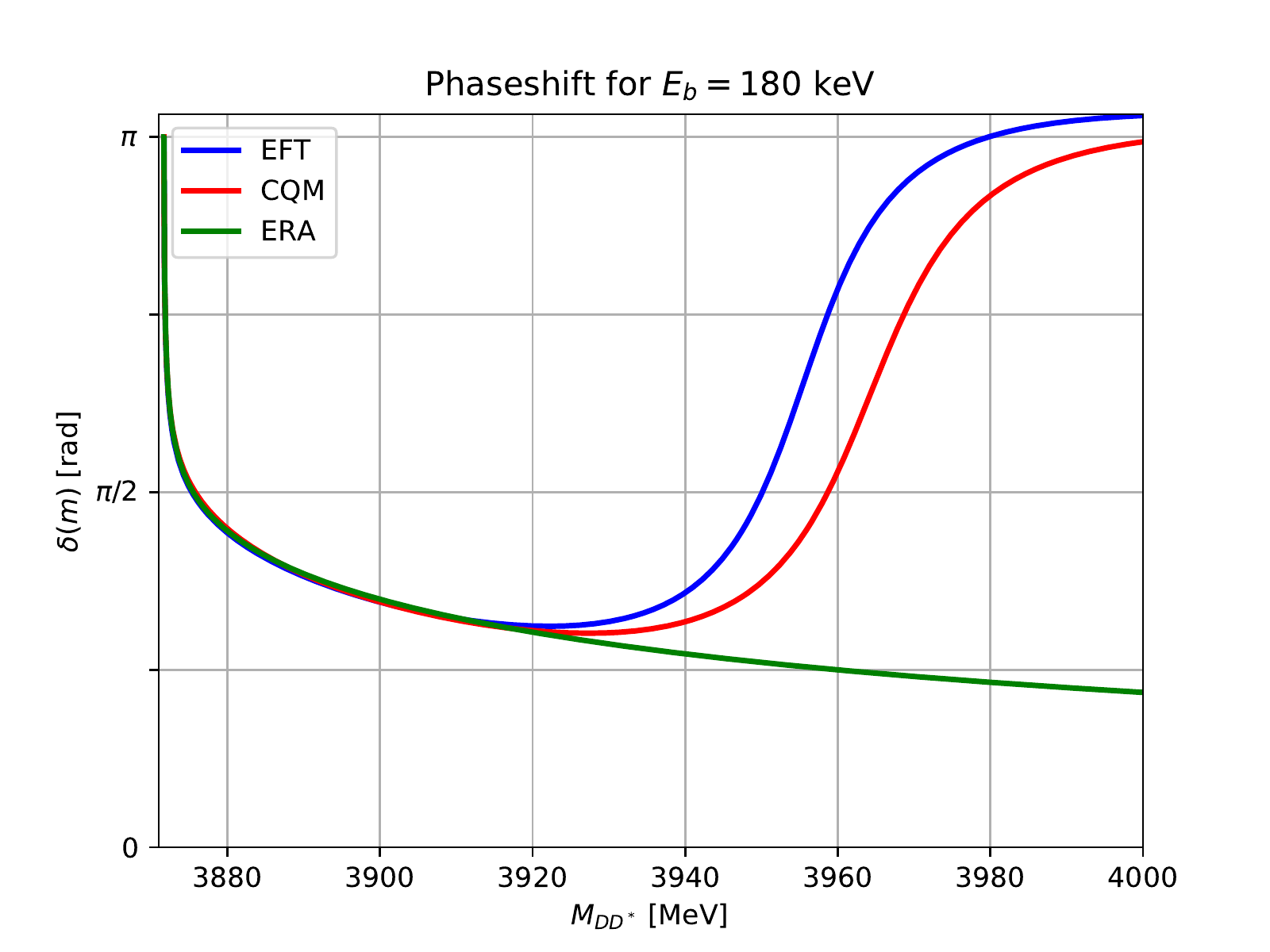}
  \includegraphics[width=.45\textwidth]{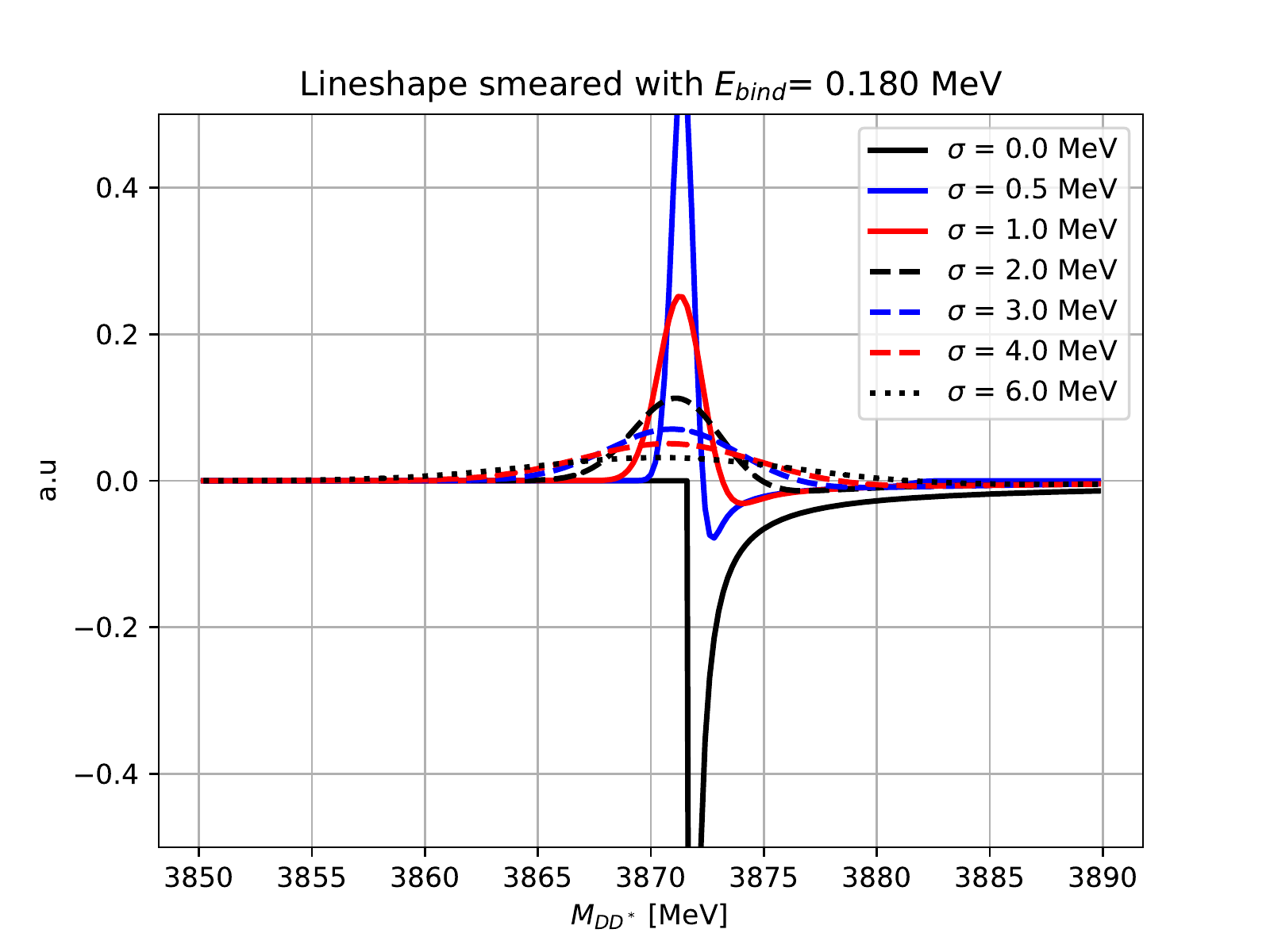}
  \end{center}
  \caption{Phase-shift for the $J^{PC}=1^{++}$ comparing the effective
    range expansion with coupled channel quark or EFT  models (left
    panel). Level density for different gaussian smearings (Right
    panel). We take $B_X=180\, {\rm KeV}$.}
\label{ps-dens}
\end{figure}

\section{Occupation number in the continuum}

The best way to illustrate the impact of the Dashen-Kane cancellation
is by looking at occupation numbers at finite temperature.  For a
single mass state the occupation number is given by
\begin{equation}\label{ec:Nocup1}
\bar n= \frac{\langle N \rangle_T}{V} =
\int \frac{d^3 k}{(2\pi)^3} \frac{g}{e^{\sqrt{k^2+m^2} /T} + \eta} = \frac{T^3}{2\pi^2} \sum_{n=1}^\infty g \frac{(-\eta)^{n+1}}{n} \left( \frac{m}{T} \right)^2 K_2(n m/T) \, , 
\end{equation}
where $\eta = \pm 1$ for Fermions or Bosons respectively and $K_2$
are modified Bessel functions. This formula
applies only to bound states, but according the quantum virial
expansion~\cite{Dashen:1969ep} the contribution of binary collisions to the
partition function should be accounted for as  a modified occupation number
in each $J^{PC} $channel, 
\begin{equation}
n(T) = \int \frac{d^3 p}{(2\pi)^3} dm \frac{g}{e^{\sqrt{p^2+m^2} /T} + \eta} \rho(m) \, ,  
\end{equation}
which
takes into account all the two particle states with their interaction
characterized by their level density in the continuum which 
can be written in terms of two-particle bound states with masses
$m_n$ and the scattering eigen phase-shifts in the channels
sharing the same $J^{PC} $ quantum numbers 
\begin{equation}
  \rho(m) = \sum_{n} \delta(m-m_n)+\frac1{\pi} \sum_{\alpha}^n \frac{d \delta_\alpha}{d m} \, , \qquad m= \sqrt{s} \,\, ({\rm CM \,\,  energy})
\end{equation}
Two interesting particular cases are worth to consider. For a (Breit-Wigner)
resonance such as $\rho \to \pi \pi$ or $\Delta \to \pi N$ one has 
  \begin{equation}
  \delta(m)=\tan^{-1} \left[ \frac{m-m_R}{\Gamma_R} \right] \to \frac1\pi \delta'(m)= \frac1\pi \frac{\Gamma_R}{(m-m_R)^2+\Gamma_R^2}
  \end{equation}
  The other case of interest is that of a weakly bound state
  close to the continuum   $d \to pn$  or $X(3872) \to D \bar D^*$ is given by the effective range expansion (ERA), 
  \begin{equation}
  p \cot \delta = - \frac1{\alpha_0}+ \frac12 r_0 p^2 + \dots \qquad m= \sqrt{p^2+M^2}  
  \end{equation}
  In Fig.~\ref{ps-dens} we show both $\delta(m)$ in the ERA
  approximation compared to coupled channel quark~\cite{Ortega:2009hj}
  and EFT models~\cite{Gamermann:2009uq,Cincioglu:2016fkm} the
  modified version of Levinson's theorem with confined channels is
  implemented~\cite{Dashen:1976cf}. We also show $\rho(m)$ for
  different gaussian smearings, where the positive and negative
  contributions illustrate the Dashen-Kane cancellation\modifs{,}{}  with the
  consequence that the corresponding occupation number in the $1^{++}$
  channel is largely reduced as compared to the elementary $X(3872)$,
  $n_{1^{++}} (T) \ll n_X (T)$ for moderate
  temperatures~\cite{Ortega:2017hpw}.  The previous formulas can be
  used in the possible production of $X(3872)$ in heavy ion collisions
  (see refs in \cite{Ortega:2017hpw}).

\section{Production rates at RHIC and LHC}

Surprisingly both the deuteron and the X(3872) have experimentally
been produced in high energy pp collisions, a
somewhat puzzling result. The underlying reason of how can such weakly
heavy particles be produced has not yet been found. Actually, the
large production cross section has been interpreted as the signature
of a compact object and possibly a tetraquark
state~\cite{Esposito:2015fsa}. However, unlike the deuteron which
leaves a track in the calorimeter, the detected $X(3872)$ is through
its $X(3872) \to \rho J/\psi,\omega J/\psi $ decay channels, which
implies a mass distribution.
Thus,  any state within the resolution
$\pm \Delta m/2$ with $J^{PC}=1^{++}$ will be recorded. Quite
generally, for an observable $O(m)$ we get
\begin{equation}
  O_{\Delta m} \equiv \int_{m-\Delta m/2}^{m+\Delta m/2}  dM \rho(M) O (M) \, .
\end{equation}
In the case $\Delta m \gg |B| \equiv |M_B-M_{\rm tr}|$
\begin{equation}
  O|_{M^B \pm \Delta m} = O(M^B) 
  + \frac1\pi \int_{M_{\rm tr}}^{M_{\rm tr}+\Delta m/2}  dM \delta_\alpha ' (M) O(M)\, .
\end{equation}
The thermodynamic arguments can be extended to $p_T$ distributions
assuming a Tsallis distribution~\cite{Ortega:2019fme}. For a pure mass state
it reads 
\begin{equation}
 \frac{d^3 N}{d^3 p}= \frac{g
   V}{(2\pi)^3}\left(1+(q-1)\frac{E(p)}{T}\right)^{-\frac{q}{q-1}} \xrightarrow{q\to 1}
 \frac{g V}{(2\pi)^3} e^{-\frac{E(p)}{T}}\, , 
\end{equation}
with $E(p) = \sqrt{m^2+p^2} $ so that the production cross
section becomes
\begin{equation}
\frac1{2\pi p_T} \frac{d\sigma (m)}{dp_T}={\cal N}\int dy\,\, E(p_T,y) 
\left[1+ \frac{q-1}{T}E(p_T,y)\right]^{\frac{q}{1-q}}
\end{equation}
with $E(p_T,y)=\sqrt{p_T^2+m^2} \cosh y $, $d^3N/(d^2 p_Tdy) \equiv
E_p d^3 N /d^3 p$ and $y=\tanh^{-1}(p_z/E_p)$ the rapidity and
${\cal N}$ the normalization.  We have found that deuteron $d$ and
$X(3872)$ accelerator production data are fully compatible with the
{\it same parameters}~\cite{Ortega:2019fme} and that in fact ${\cal
  N}_X \sim {\cal N}_d$,  so that there is nothing more special about
the $X$ than the deuteron, and so its large production rate at high
$p_T$. 

\begin{figure}[h]
 \includegraphics[width=.45\textwidth]{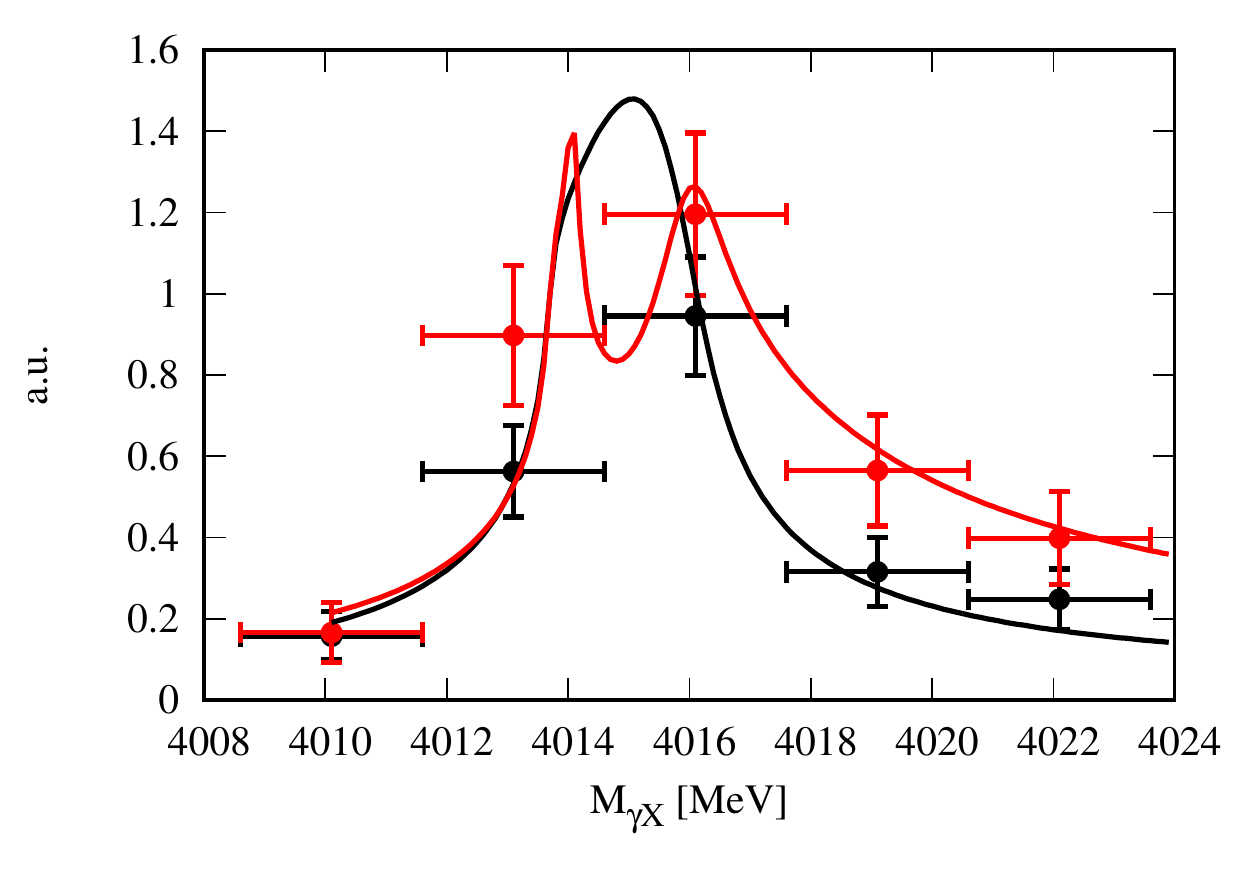}
 \includegraphics[width=.45\textwidth]{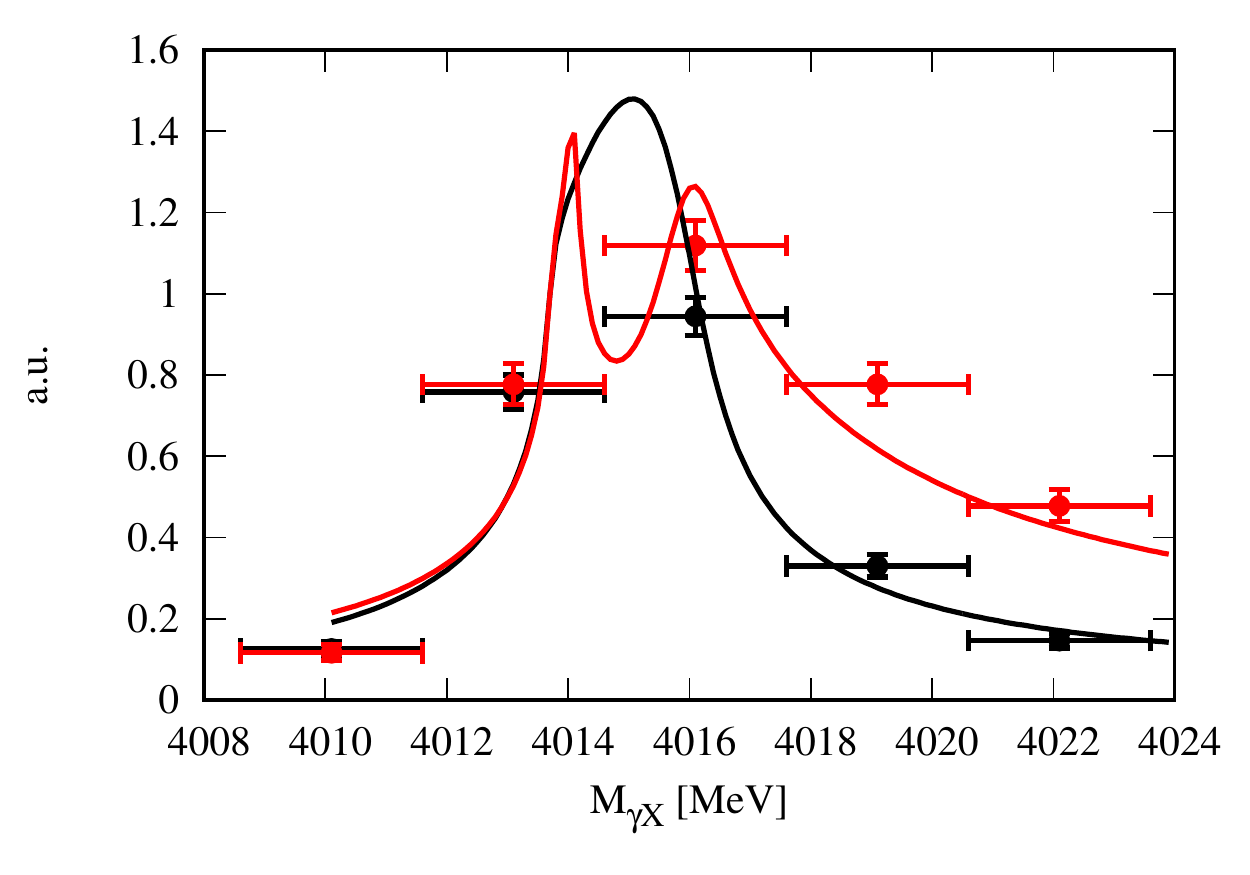}
 \caption{\label{fig6} Binned smeared lineshapes of states for
   $\sigma=1$ MeV, $E_w=3$ MeV and $\Delta M=20$ MeV for $E_b=180$ keV
   (black points) and $E_b=-180$ keV (red points).  Full lines show
   the no-binned smeared lineshape (same color code). The results are
   done with $N=100$ (left), $N=1000$ (right). }
\end{figure}

\section{Detection methods and accurate mass measurements}

Recent proposals to make accurate meassurements of mass based on
triangle singularities~\cite{Guo:2019qcn,Braaten:2019gfj} $e^+ e^- \to
X(3872) \gamma$ have been made since the lineshape is very sensitive
to the binding energy, assuming no nearby continuum effects inherent
in the detection problem. If we fold the level density $\rho(m)$ in
the $J^{PC}=1^{++}$ channel with the detector efficiency function
\begin{equation}
 R_\sigma(m,M)=\frac{1}{\sqrt{2\pi}\sigma} e^{-\frac{(m-M)^2}{2\sigma^2}}
\end{equation}
we get corresponding smeared lineshape profiles (see
Fig.~\ref{ps-dens} right panel). A Monte Carlo simulation of the
effects in the smeared profiles for different samples with $N=100$ and
$N=1000$ is shown in Fig.~\ref{fig6}.  As we see, the large
differences advocated in~\cite{Guo:2019qcn,Braaten:2019gfj},
corresponding to the $\sigma \to 0$ case, are somewhat blurred, and
set a standard on the necessary statistics discriminating different
signals.  More details will be presented elsewhere.

\section{Conclusions}

Regardless of its detailed  molecular or quark-antiquark or mixed nature and 
composite structure, the weakly bound $X(3782)$ can be best regarded
as a mass distribution as long as the operating resolution is much
larger than the binding energy. The reason is that the signal of
$X(3872)$ is by its decay products recorded within the detector
resolution, and the mass distribution exhibits a cancellation between
the bound state and the continuum. This is most clearly seen by
analyzing the occupation number at finite temperature and in high
energy proton-proton collisions which show clear departures from the
elementary limit. This cancellation is also of relevance in possible
precision measurements based on a lineshape profile sensitive to the
numerical value binding energy.


\end{document}